\pdfoutput=1
\documentclass[fleqn,usenatbib,usedcolumn]{mnras}

\usepackage{graphicx,amssymb,hyperref}
\usepackage{multirow}
\usepackage{color}

\usepackage[fleqn]{amsmath}
\usepackage{abbrev}
\usepackage{siunitx}
\usepackage{comment} 
\usepackage{gensymb}
\usepackage{booktabs,tabularx}
\usepackage{flushend}

\usepackage{nicefrac}
\usepackage{units}
\usepackage{verbatim}

\def\mnras{MNRAS}
\def\apj{ApJ}

\def\simlt{\lower.5ex\hbox{$\; \buildrel < \over \sim \;$}}
\def\simgt{\lower.5ex\hbox{$\; \buildrel > \over \sim \;$}}
\def\etal{{\it et al.}}

\def\D{\mathrm{d}}

\def\kpc{\mathrm{\, kpc}}
\def\mpc{\mathrm{\, Mpc}}
\def\msun{\mathrm{\, M_\odot}}
\def\kms{\mathrm{\, km \, s^{-1}}}
\def\cmsg{\, \mathrm{cm^2 \, g^{-1}}}
\def\gyr{ \, \mathrm{Gyr}}
\def\myr{ \, \mathrm{Myr}}
\def\sigT{\sigma_{T}}
\def\sigTt{\sigma_{\tilde{T}}}
\def\sTt{modified momentum-transfer cross-section}

\newcommand{\be}{\begin{equation}}
\newcommand{\ee}{\end{equation}}
\newcommand{\ba}{\begin{eqnarray}}
\newcommand{\ea}{\end{eqnarray}}

\DeclareGraphicsExtensions{.pdf,.png}

\title[Anisotropic DM Scattering]{Cosmic particle colliders: simulations of self-interacting dark matter with anisotropic scattering}

\author[A.\ Robertson \etal]{Andrew Robertson\thanks{e-mail: {\tt andrew.robertson@durham.ac.uk}}, 
Richard Massey and
Vincent Eke\\Institute for Computational Cosmology, Durham University, South Road, Durham DH1 3LE, UK\\
}

\begin{document}

\maketitle

\label{firstpage}

\begin{abstract}
We investigate how self-interacting dark matter (SIDM) with anisotropic scattering affects the evolution of isolated dark matter haloes as well as systems with two colliding haloes. For isolated haloes, we find that the evolution can be adequately captured by treating the scattering as isotropic, as long as the isotropic cross-section is appropriately matched to the underlying anisotropic model. We find that this matching should not be done using the momentum transfer cross-section, as has been done previously. Matching should instead be performed via a modified momentum transfer cross-section that takes into account that dark matter particles can be relabelled after they scatter, without altering the dynamics. However, using cross-sections that are matched to give the same behaviour in isolated haloes, we find that treating dark matter scattering as isotropic underpredicts the effects of anisotropic dark matter scattering when haloes collide. In particular, the DM-galaxy offset induced by SIDM in colliding galaxy clusters is larger when we simulate the underlying particle model, than if we use a matched isotropic model. On the other hand, well motivated particle models with anisotropic scattering typically have cross-sections with a strong velocity dependence, and we discover a previously unrecognised effect that suppresses DM-galaxy offsets in colliding clusters making it hard for these systems to provide competitive constraints on such particle models. 
\end{abstract}


\section{Introduction}

Despite mounting astrophysical evidence for the existence of dark matter (DM) as the dominant matter component of the Universe \citep[e.g.][]{2016A&A...594A..13P}, its nature remains a mystery. It is usually assumed to be a cold and collisionless particle (CDM), for which the predictions for the large scale structure of the Universe provide a striking match to what is observed \citep{2016MNRAS.460.1173R}. However, there are possible modifications to CDM that preserve this success on large scales, while altering DM's behaviour in collapsed objects. One possible modification is to allow DM particles to scatter elastically at rates that are astrophysically interesting. While some of the most popular DM candidates (for instance supersymmetric neutralinos) interact only through gravity and the weak force and behave as collisionless particles during structure formation, a high rate of DM scattering from so-called self-interacting dark matter (SIDM) would be possible if the DM lives in a rich dark sector with a new dark force.

The presence of such a dark force would have significant implications for both particle physics and astrophysics. Large self-interactions would rule out some popular DM candidates such as axions \citep{2009NJPh...11j5008D}, and would change cosmological structure formation on small scales. These changes to small-scale structure are appealing as they could resolve discrepancies between the results of $N$-body simulations with CDM and observations of dwarf galaxies \citep[for a review see][]{Weinberg02022015}.

The tightest constraints on DM's self-interaction cross-section have come from galaxy cluster scales \citep{2002ApJ...564...60M,Randall:2008hs,Rocha:2013bo,2013MNRAS.430..105P,2015Sci...347.1462H,2016arXiv160808630K}, while the astrophysical motivation for SIDM predominantly comes from dwarf galaxies \citep{2012MNRAS.423.3740V,2013MNRAS.431L..20Z,2015MNRAS.453...29E,2016MNRAS.460.1399V}. The typical velocities of DM particles within galaxy clusters are of the order of $1000 \kms$, while in dwarf galaxies they can be 10 to 100 times lower. Given that it is common for scattering cross-sections to have a strong velocity dependence (such as $\sigma \propto v^{-4}$ in the case of Rutherford scattering) and that this is true also of many particle physics based models for SIDM \citep{2009PhRvD..79b3519A,2010PhRvL.104o1301F,2010PhRvD..81h3522B}, it is not unreasonable for the cross-section in dwarf galaxies to be orders of magnitude larger than in galaxy clusters. This has led to such particle candidates being simulated \citep{2012MNRAS.423.3740V,2013MNRAS.431L..20Z,2013MNRAS.430.1722V,2014MNRAS.444.3684V} in a bid to alleviate tensions on small scales, while evading constraints that come from larger scales.

The parameters governing such velocity-dependent cross-sections (DM mass, mediator mass and coupling strength) can in principle be constrained by estimating the cross-section for DM--DM scattering at different velocities. \citet{2016PhRvL.116d1302K} recently estimated the DM mass and dark photon mass, assuming that the inferred core sizes in observed galaxy clusters \citep{2013ApJ...765...25N}, low surface brightness galaxies \citep{2008ApJ...676..920K} and dwarf galaxies \citep{2011AJ....141..193O} are due to SIDM, and using a coupling strength equal to the electromagnetic fine structure constant $\alpha' = \alpha \approx 1/137$.

What is often ignored when simulating these velocity-dependent cross-sections is that the scattering is usually anisotropic. This is because the velocity dependence results from a term in the scattering cross-section that depends on the exchanged momentum, which depends on both the collision velocity and the scattering angle. This angular dependence has not been included in previous simulations, which have instead simulated the scattering as isotropic but with a cross-section modified such that the effects of DM scattering should be similar to what would result from a faithful simulation using the underlying particle interaction. This has been done by matching the momentum transfer cross-section as a function of collision velocity, $\sigT(v)$,  between the true particle interaction and that used in the simulations. While this may work well when the DM velocity distribution is close to isotropic, \citet[][hereafter K14]{2014MNRAS.437.2865K} found that for the case of colliding galaxy clusters the momentum transfer cross-section is insufficient to fully characterise the effects of DM scattering. This is not surprising; when galaxy clusters collide there is a strongly preferred direction along which DM particles collide and the angular distribution of scattered DM becomes important.

The goal of this paper is to explore the effects of anisotropic DM scattering, by simulating scattering processes faithfully to their underlying particle physics models. We introduce a new code that can simulate scattering with a general differential cross-section and explore how the results from anisotropic scattering compare with the case of isotropic scattering for the evolution of an isolated DM halo, as well as in a galaxy cluster collision. By also simulating these systems using an appropriately matched isotropic cross-section, as has been done in the past, we can test the validity of this approximate scheme. 

This work is organised as follows. In \S\ref{sec:Ang_Dep_Scatt} we discuss the physics of anisotropic scattering and introduce two examples of anisotropic scattering cross-sections. We discuss the implementation of angular dependent scattering within an $N$-body code in \S\ref{sec:scattering_in_code} and then use this code to investigate how core-formation in an isolated DM halo depends on the angular-dependence of the scattering cross-section. In \S\ref{sec:offsets_in_BC} we show the results of simulations of the Bullet Cluster run with anisotropic scattering, and contrast this system that has a clear directionality to the case of an isolated halo. Finally, we summarise our results in \S\ref{sec:conclusions}.

\section{Angular Dependent Scattering}
\label{sec:Ang_Dep_Scatt}

The key particle physics quantity for a study of the astrophysical effects of SIDM is the differential cross-section, $\D \sigma / \D \Omega$.
This quantifies the rate at which particles are scattered into different patches of solid angle, and can vary as a function of the collision velocity, $v$.

In this section we briefly describe the particle physics that leads to angular-dependent scattering, and then describe different integrated cross-sections, which condense a scattering probability that varies with angle into a single number. We end the section by introducing two different anisotropic cross-sections, which we later use in our simulations.

When dealing with particle scattering we assume that DM particles are indistinguishable and work in the centre of momentum frame of the two interacting particles, with the velocities of the DM particles defined to be $\pm v/2$ in the $z$-direction. To conserve both energy and momentum, both particles leave the collision with a velocity of $v/2$ at a polar angle $\theta$ with respect to their incoming direction, where $\theta$ must be the same for the two particles. Assuming that the scattering potential is spherically symmetric, the differential cross-section is independent of the azimuthal angle $\phi$. The DM particle mass is $m_\chi$ and the DM scattering is mediated by a particle with mass $m_\phi$, with a coupling strength (analogous to the fine structure constant) $\alpha_\chi$.

\subsection{Particle physics of angular dependent scattering}
\label{sect:part_phys_ang_dep}

As mentioned in the Introduction, most efforts to simulate SIDM have treated the DM scattering as isotropic. This isotropic scattering, also commonly referred to as `contact interactions', `hard sphere scattering' and `billiard ball scattering'\footnote{Readers may be interested to note that the cross-section per unit mass of a World Pool-Billiard Association ball is $0.64 \cmsg$, of similar magnitude to commonly studied SIDM cross-sections, though it is unclear how resilient pool balls would be to $\sim 1000 \kms$ collisions.} results from an interaction with a massive mediator, leading to a short range force. For DM particles with a relative velocity $v$, scattering will be isotropic when the mediator mass is much heavier than the DM particle momenta, $c \, m_\phi \gg v \, m_\chi$. When this is not the case, the interaction cross-section will typically depend on the exchanged momentum, which increases with an increased collision velocity or an increased scattering angle, leading to velocity-dependent anisotropic scattering. This second case, with long range interactions due to a light or massless mediator, arises in models of mirror DM \citep{1983SvA....27..371B,1996PhLB..375...26B,2004IJMPD..13.2161F} and atomic DM \citep{2012PhRvD..85j1302C,2013PhRvD..87j3515C}, as well as some other hidden sector DM models \citep{2009JCAP...07..004F,2015PhRvD..91b3512F,2016PhRvD..94l3017B}.

\subsection{Integrated cross-sections}

Given an azimuthally-symmetric differential cross-section, $\D \sigma / \D \Omega$, the total cross-section
\begin{equation}
\label{eq:tot_cross_sect}
\sigma \equiv \int \frac{\D \sigma}{\D \Omega} \D \Omega = 2 \pi \int \frac{\D \sigma}{\D \Omega} \sin \theta \, \D \theta.
\end{equation}
While this is the relevant quantity when considering the rate at which particles interact, it does not fully describe the consequences of these interactions, as the effect of scattering by a large angle (and so transferring a large amount of momentum between the two particles) is greater than the effect of scattering by a small angle.

A useful concept when comparing the macroscopic consequences of particle interactions with different angular dependencies for the differential cross-section is the momentum-transfer cross-section. For a scattering angle of $\theta$, the momentum transfer along the direction of the collision is
\begin{equation}
\label{eq:delta_px}
\Delta p_z = p (1 - \cos \theta),
\end{equation}
where $p$ is the magnitude of each of the incoming particles' momenta in the centre of momentum frame. We therefore define the momentum-transfer cross section as
\begin{equation}
\label{eq:sigT}
\sigma_T \equiv \int (1 - \cos \theta) \frac{\D \sigma}{\D \Omega} \D \Omega.
\end{equation}
This is similar to the definition of $\sigma$, except that interactions that lead to a large amount of momentum transfer contribute more, while those that transfer little momentum are down-weighted. For the case of isotropic scattering, where $\frac{\D \sigma}{\D \Omega}=\frac{\sigma}{4 \pi}$ is independent of angle, the momentum-transfer cross-section and the cross-section are equal, i.e. $\sigma_T = \sigma$.

K14 point out that this definition of $\sigma_T$ overestimates the momentum transfer due to scattering with $\theta > \pi / 2$, as in these cases the particles, which we assume to be identical, could be relabelled in such a way that they had scattered with $\theta < \pi / 2$. If we weight scatters by the amount of momentum transfer, but relabel particles if they scatter by an angle greater than $\pi / 2$, then we get the integrated cross-section
\begin{equation}
\label{eq:sigTt}
\sigma_{\tilde{T}} \equiv \int_{\theta=0}^{\pi/2} (1 - \cos \theta) \frac{\D \sigma}{\D \Omega} \D \Omega + \int_{\theta=\pi/2}^{\pi} (1 + \cos \theta) \frac{\D \sigma}{\D \Omega} \D \Omega,
\end{equation}
which we call the \sTt. For isotropic scattering $\sigma_{\tilde{T}} = \sigma / 2$, while for cross-sections with a negligible amount of large-angle scattering $\sigma_{\tilde{T}} \approx \sigma_T$.\footnote{For most cross-sections $\sigma_{\tilde{T}}$ has a similar value to the viscosity (or conductivity) cross section $\sigma_V \equiv \int  \sin^2 \theta \frac{\D \sigma}{\D \Omega} \D \Omega$ advocated by \citet{2013PhRvD..87k5007T}, \citet{2014PhRvD..89d3514C} and \citet{2016PhRvD..94l3017B} for reasons similar to those for introducing $\sigma_{\tilde{T}}$. We also note however, that \citet{2016arXiv161004611A} argue against using such a procedure for the cross-sections they consider.}

While integrated cross-sections such as $\sigma_T$ and $\sigma_{\tilde{T}}$ do not fully describe a scattering process, they are useful as a way to compare different scattering cross-sections, and have been used in cosmological simulations of SIDM. The reason for this is computational efficiency. For anisotropic cross-sections, where the vast majority of scattering events involve a low momentum transfer (such as scattering from a Coulomb potential), there can be a large number of interactions, each having very little effect. A less computationally intensive way to simulate a similar effect is to simulate the scattering as isotropic, where most scattering events involve a large amount of momentum transfer, but with a total cross-section scaled down so that the rate of momentum transfer matches the momentum transfer expected from the underlying particle physics model for the DM. This reduces the number of interactions that need to be calculated, while attempting to adequately capture the effects of particle scattering.

\subsection{A velocity-independent, anisotropic cross-section}

In \S\ref{sect:part_phys_ang_dep} we discussed that anisotropic scattering usually occurs when the cross-section is also velocity-dependent. However, studying an anisotropic cross-section without velocity dependence is useful to gain intuition for what might happen with more complicated cross-sections, and if realised in nature could have interesting effects in merging galaxy clusters (K14). As an example of such a cross-section we use:
\begin{equation}
\label{eq:KVI_cross_sect}
\frac{\D \sigma}{\D \Omega} = \frac{\alpha^2}{2 m_\chi^2} \frac{1 + \cos^2 \theta}{1 - \cos^2 \theta}
\end{equation}
for which both $\sigma$ and $\sigma_T$ diverge and which we call Kahlhoefer velocity-independent (KVI). In the case of the $\sigma_T$ divergence, this is because of the divergence in the differential cross-section as $\theta \to \pi$. While scattering by $\sim \pi$ leads to a significant amount of momentum transfer between the two particles, it leaves the system relatively unchanged, as two identical particles just swap velocities with each other. For this reason, $\sigma_{\tilde{T}}$ is a more sensible choice to describe the scattering. For this differential cross-section
\begin{equation}
\label{eq:KVI_sigmaT}
\sigma_{\tilde{T}} = \frac{\pi \alpha^2}{m_\chi^2} \left (\ln 16 - 1 \right).
\end{equation}

For the KVI cross-section, the divergence in the differential cross-section as $\theta \to 0$ and $\theta \to \pi$, means that we cannot simulate the cross-section completely faithfully. However, the \sTt \, is finite for this differential cross-section because the divergence in cross-section at low angles is accompanied by a suitably rapid decline in the effectiveness of these scatters to transfer momentum. This means that for a small $\theta_\mathrm{min}$, one should expect that ignoring scattering with $\theta < \theta_\mathrm{min}$ and $\theta > \pi - \theta_\mathrm{min}$, should lead to negligible changes to the effect of this cross-section.

Introducing a cut-off, such that the differential cross-section follows equation~\eqref{eq:KVI_cross_sect} for $\theta_\mathrm{min} < \theta < \pi - \theta_\mathrm{min}$ and is 0 outside of this, the cross-section is then finite,
\begin{equation}
\label{eq:KVI_thetamin_sigma}
\sigma(\theta_\mathrm{min}) = \frac{2 \pi \alpha^2}{m_\chi^2} \left\{ \ln\left(\frac{1 + \cos \theta_\mathrm{min} }{1 - \cos \theta_\mathrm{min}}\right) - \cos \theta_\mathrm{min} \right\}.
\end{equation}
The \sTt \, becomes
\begin{equation}
\label{eq:KVI_thetamin_sigmaT}
\begin{split}
\sigma_{\tilde{T}}(\theta_\mathrm{min}) = \frac{\pi \alpha^2}{m_\chi^2}  \left\{ \right. \cos^2 \theta_\mathrm{min} &- 2\cos \theta_\mathrm{min}  \\
	&+ \left. 4 \ln (1 + \cos \theta_\mathrm{min}) \right\}
\end{split}
\end{equation}
which can be compared with equation~\eqref{eq:KVI_sigmaT} to see how much momentum-transfer we expect to miss by introducing $\theta_\mathrm{min}$. In particular, for $\theta_\mathrm{min} \ll 1$
\begin{equation}
\label{eq:KVI_thetamin_sigmaT_ll1}
\sigma_{\tilde{T}} - \sigma_{\tilde{T}}(\theta_\mathrm{min}) = \frac{\pi \alpha^2}{m_\chi^2} \left\{ \theta_\mathrm{min}^2 + \mathcal{O}(\theta_\mathrm{min}^4)\right\}.
\end{equation}
As an example, with $\theta_\mathrm{min} = 0.1$ we only lose 0.6\% of $\sigma_{\tilde{T}}$.

\subsection{Yukawa-potential SIDM}
\label{sect:Yukawa_SIDM}

A general result for scattering mediated by a massive mediator particle is that it is equivalent to having a Yukawa potential. \citet{2011PhRvL.106q1302L} noted that such a cross-section would display interesting astrophysical signatures because the rate of scattering peaks at a particular pairwise velocity, falling at smaller or larger velocities. This could lead to scattering being important in DM haloes of a particular mass (and so a particular velocity dispersion), while being negligible in the more massive haloes that have thus far provided the tightest constraints on the DM cross-section. Simulations including such a model for DM scattering have been performed, but have simulated the scattering as isotropic, using the momentum-transfer cross-section of the underlying particle physics model, as the cross-section for isotropic scattering \citep{2012MNRAS.423.3740V,2013MNRAS.430.1722V,2013MNRAS.431L..20Z,2014MNRAS.444.3684V}. We call this method of simulating DM models with anisotropic cross-sections, $\sigma_T$-match.

There is no analytical form for the differential scattering cross-section due to a Yukawa potential, but by using the Born-approximation \citep{1990fnp..book.....J}, valid when the scattering potential can be treated as a small perturbation, we can find an analytical form that approximates the true differential cross-section. For an interaction potential given by
\begin{equation}
\label{eq:yukawa_V}
V(r) = - \frac{\alpha_\chi e^{-m_\phi r}}{m_\phi r}, 
\end{equation}
the differential cross-section assuming the Born approximation is \citep{2010PhLB..692...70I}
\begin{equation}
\label{eq:yukawa_differential_cross-sect}
\frac{\D \sigma}{\D \Omega} = \frac{\alpha_\chi^2}{m_\chi^2 \left(m_\phi^2/m_\chi^2  + v^2 \sin^2 \frac{\theta}{2} \right)^2 }, 
\end{equation}
where we have used natural units with $\hbar = c = 1$. This can be re-written as 
\begin{equation}
\label{eq:yukawa_differential_cross-sect_alt}
\frac{\D \sigma}{\D \Omega} = \frac{\sigma_0}{4 \pi (1 + \frac{v^2}{w^2} \sin^2 \frac{\theta}{2})^2 },
\end{equation}
where $w= m_\phi c / m_\chi$ is a characteristic velocity, below which the scattering is roughly isotropic with $\sigma \approx \sigma_0$. At higher velocities, the scattering has an angular dependence that tends to that from scattering with a Coulomb potential, with a cross-section that decreases with increasing velocity.

From the differential cross-section we can calculate the integrated cross-sections
\begin{equation}
\label{eq:yukawa_sigma}
\sigma = \frac{\sigma_0}{1 + \frac{v^2}{w^2}},
\end{equation}
\begin{equation}
\label{eq:yukawa_sigmaT}
\sigma_T = \sigma_0 \frac{2 w^4}{v^4} \left\{ \ln \left( 1 + \frac{v^2}{w^2} \right) - \frac{v^2}{w^2+v^2} \right\},
\end{equation}
and
\begin{equation}
\label{eq:yukawa_sigmaTt}
\sigma_{\tilde{T}} = \sigma_0 \frac{2 w^4}{v^4} \left\{ 2 \ln \left(1 + \frac{v^2}{2 w^2} \right) - \ln \left(1 + \frac{v^2}{w^2} \right) \right\}.
\end{equation}

We note that at low velocities the scattering is non-perturbative and the Born approximation is no longer valid. This is the reason why the behaviour of $\sigT$ in equation~\eqref{eq:yukawa_sigmaT} differs from that found by numerically solving for orbits in a classical Yukawa potential \citep{2004ITPS...32..555K} at low velocities. The Born approximation result tends towards isotropic scattering with a velocity-independent cross-section, while the results of the full calculation have a cross-section that logarithmically increases towards low velocities. We therefore expect slightly different results compared with a full calculation of scattering through a Yukawa potential. However, using the Born approximation is useful as it gives us an analytical differential cross-section that we can faithfully simulate, allowing us to test the procedure of using isotropic scattering to capture the effects of a more complicated scattering process. With a known differential cross-section, we can simulate the scattering in a fully consistent manner, and compare that with simulating it with a suitably matched $\sigma(v)$ and isotropic scattering.

\section{Implementing DM scattering}
\label{sec:scattering_in_code}

Our implementation of DM scattering follows that of \citet{Robertson11022017} who implemented isotropic scattering of DM in the \textsc{gadget}-3 Tree-PM $N$-body code, which is an updated version of the publicly available \textsc{gadget}-2 code \citep{Springel:2005cz}. At each time-step particles search for neighbouring particles, separated by a distance less than the search radius $h$, and scatter from each of their neighbouring particles with a probability
\begin{equation}
\label{eq:P_ij}
P_\mathrm{scat}= \frac{\sigma v \, \Delta t}{\frac{4\pi}{3}h^{3}},
\end{equation}
where $\sigma$ is the DM scattering cross-section, $v$ is the relative velocity of the two particles, and $\Delta t$ is the size of the time-step. $h$ is a numerical parameter that we keep fixed for all particles, with a size similar to the gravitational softening length $\epsilon$.

All simulations discussed in Sections~\ref{sect:core_growth} and \ref{KVI Bullet Cluster} used $h = 2 \epsilon = 5.6 \kpc$, while in Section~\ref{sect:yukawa_BC} larger $h$ (up to $4 \epsilon$) were used. In general, a smaller $h$ is better, because scattering is then more local and the scattering rate resolves small-scale density peaks. However, with smaller $h$, fewer neighbour particles are found, so the probability of scattering from those particles must increase to achieve the correct rate of scattering. If $h$ is too small, such that $P_\mathrm{scat} > 1$, then the rate of scattering will no longer be correctly calculated. To avoid this, small time-steps can be used, but this makes the simulations computationally expensive. For this reason, we use larger $h$ for our simulations that have large cross-sections, keeping the maximum value of $P_\mathrm{scat}$ below 0.1 in all cases. A more detailed discussion about the choice of $h$ can be found in \citet{Robertson11022017}.

We extend our code by allowing for the cross-section to depend on velocity, and also for the polar scattering angle, $\theta$, to be drawn from an anisotropic probability distribution. The velocity-dependence is easily achieved by allowing $\sigma$ in equation~\eqref{eq:P_ij} to be a function of velocity, while the angular-dependence is realised by generating tables of scattering angles drawn from the relevant probability distribution. A full discussion of our implementation, along with a test of scattering from a general differential cross-section, is contained in Appendix~\ref{App:scattering_implementation}. 

\section{Core growth in isolated haloes}
\label{sect:core_growth}

In an isolated halo with an isotropic velocity distribution, there is no preferred direction for particle scattering, and a suitably matched isotropic cross-section may be able to mimic the effects of an anisotropic one. To test the efficacy of the $\sigma_T$-match procedure defined in \S\ref{sect:Yukawa_SIDM}, we investigate the rate at which cores form in an isolated Hernquist profile \citep{Hernquist:1990tq} with anisotropic scattering and compare to isotropic scattering. We simulate a halo with parameters corresponding to the bullet halo in the Bullet Cluster, namely $M = \num{2.46e14} \msun$ and $a = 279 \kpc$ \citep{Robertson11022017}. 

\subsection{Determining core sizes}

We find the core size, $r_\mathrm{core}$, by fitting a cored Hernquist profile
\begin{equation}
\label{eq:cored_Hernquist}
\rho(r) = \frac{M}{2 \pi} \frac{a}{(r^\beta+r_\mathrm{core}^\beta)^{1/\beta}} \frac{1}{(r+a)^3}
\end{equation}
to the radial density distribution. During the fitting procedure we allow $M$, $a$, and $r_\mathrm{core}$ to vary, while holding $\beta = 4$ fixed. We measure the core evolution in terms of a dimensionless time, $T/T_\mathrm{dyn}$, where the dynamical time follows the definition in \citet{2000ApJ...543..514K},
\begin{equation}
\label{eq:T_dyn}
T_\mathrm{dyn}= 4 \pi \sqrt{\frac{a^3}{G M}},
\end{equation}
is $1.8 \gyr$ for our chosen Hernquist profile. We run simulations with different cross-sections, which again are defined to be dimensionless
\begin{equation}
\label{eq:dimensionless_sigma}
\hat{\sigma}= \frac{\sigma}{m} \frac{M}{a^2},
\end{equation}
such that $\hat{\sigma} = 1$ corresponds to $\sigma/m \approx1.5 \cmsg$ for our simulated halo.

\subsection{KVI scattering}
\label{sect:KVI_Hernquist}

\begin{figure}
        \centering
        \includegraphics[width=\columnwidth]{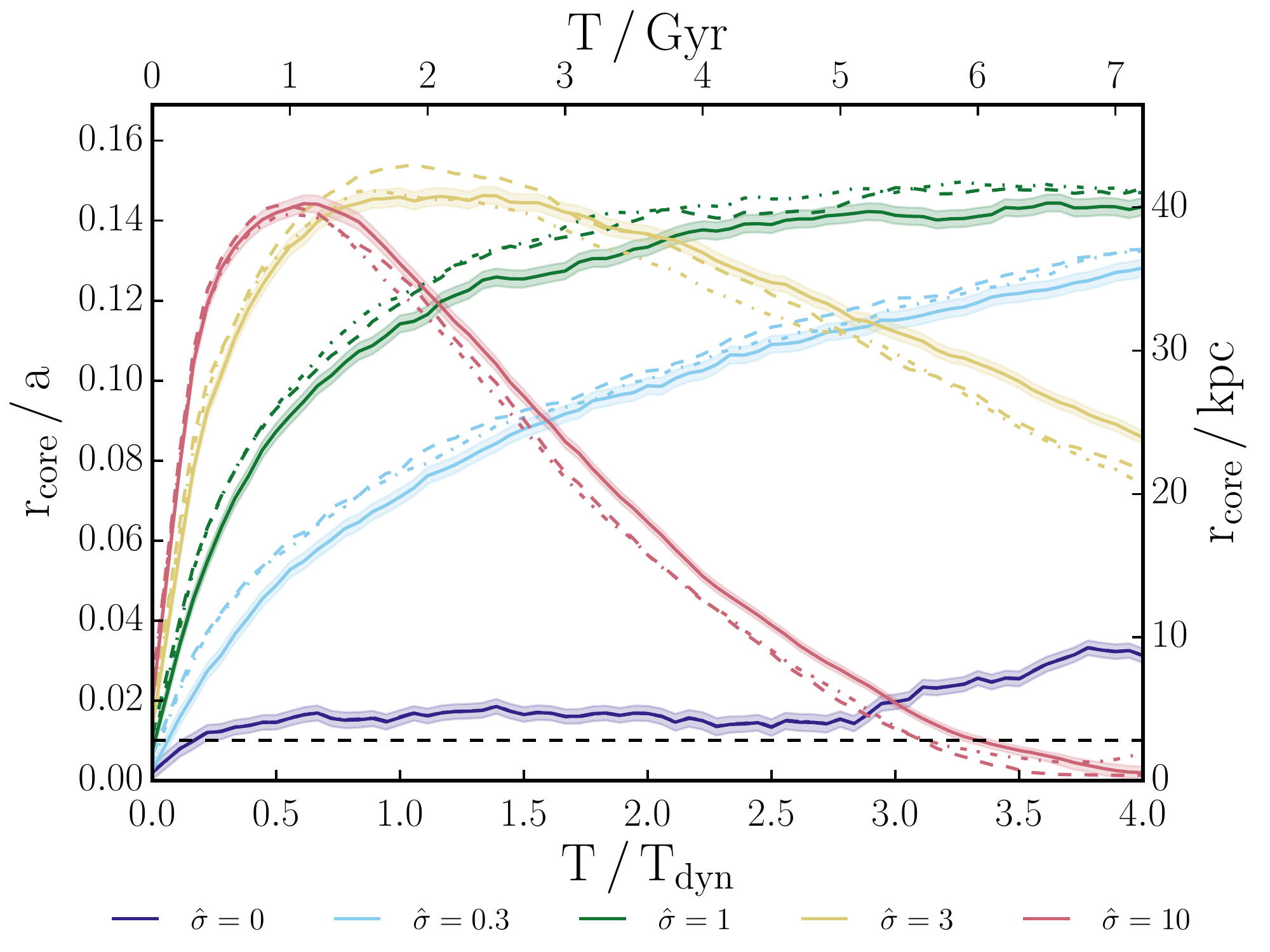}
	\caption{The evolution of core size for an isolated Hernquist profile, evolved with different SIDM scattering cross-sections. The solid lines are for isotropic DM scattering, while the dashed and dot-dashed lines are for anisotropic KVI scattering with $\theta_\mathrm{min}=0.025$ and 0.1 respectively, matched to the isotropic cross-sections using $\sigTt$. In the case of the dashed line, there is between 4 and 5 times as much scattering as for the equivalent solid line, but due to the angular-dependence of those interactions the resulting evolution is similar. The shaded regions around the solid lines show the $1 \sigma$ error on $r_\mathrm{core}$ and the horizontal dashed line shows the size of the gravitational softening length.}
	\label{fig:core_radius_evolution}
\end{figure}

With a KVI cross-section we cannot use $\sigma_T$-match, as the momentum transfer cross-section diverges. Instead we match $\sigTt$ between KVI scattering and isotropic scattering. In Fig.~\ref{fig:core_radius_evolution} we demonstrate the effectiveness of $\sigTt$-match, showing that when the KVI cross-section is normalised such that it has the same $\sigTt$ as a particular isotropic cross-section, then the evolution of the core size is close to that for the matched isotropic cross-section. We plot the results using $\theta_\mathrm{min}=0.025$ and $\theta_\mathrm{min}=0.1$, which respectively correspond to 0.04\% and 0.6\% of $\sigTt$ being truncated. For $\theta_\mathrm{min}=0.025 (0.1)$ the total cross-section, $\sigma$, is 4.4(2.8) times larger than for the $\sigTt$-matched isotropic cross-section.

The similarity between the $r_\mathrm{core}$ evolution with isotropic scattering and with $\sigTt$-matched anisotropic scattering suggests that at least in locally isotropic systems $\sigTt$ is a useful way to characterise DM scattering. Using only isotropic scattering and a calculation of $\sigTt$ (and not the full, underlying differential cross-section that leads to it) we can predict how a system would evolve with anisotropic scattering.

\subsection{Yukawa-potential scattering}

\begin{figure*}
        \centering
	\begin{tabular}{ll}
  	\includegraphics[width=0.5\textwidth]{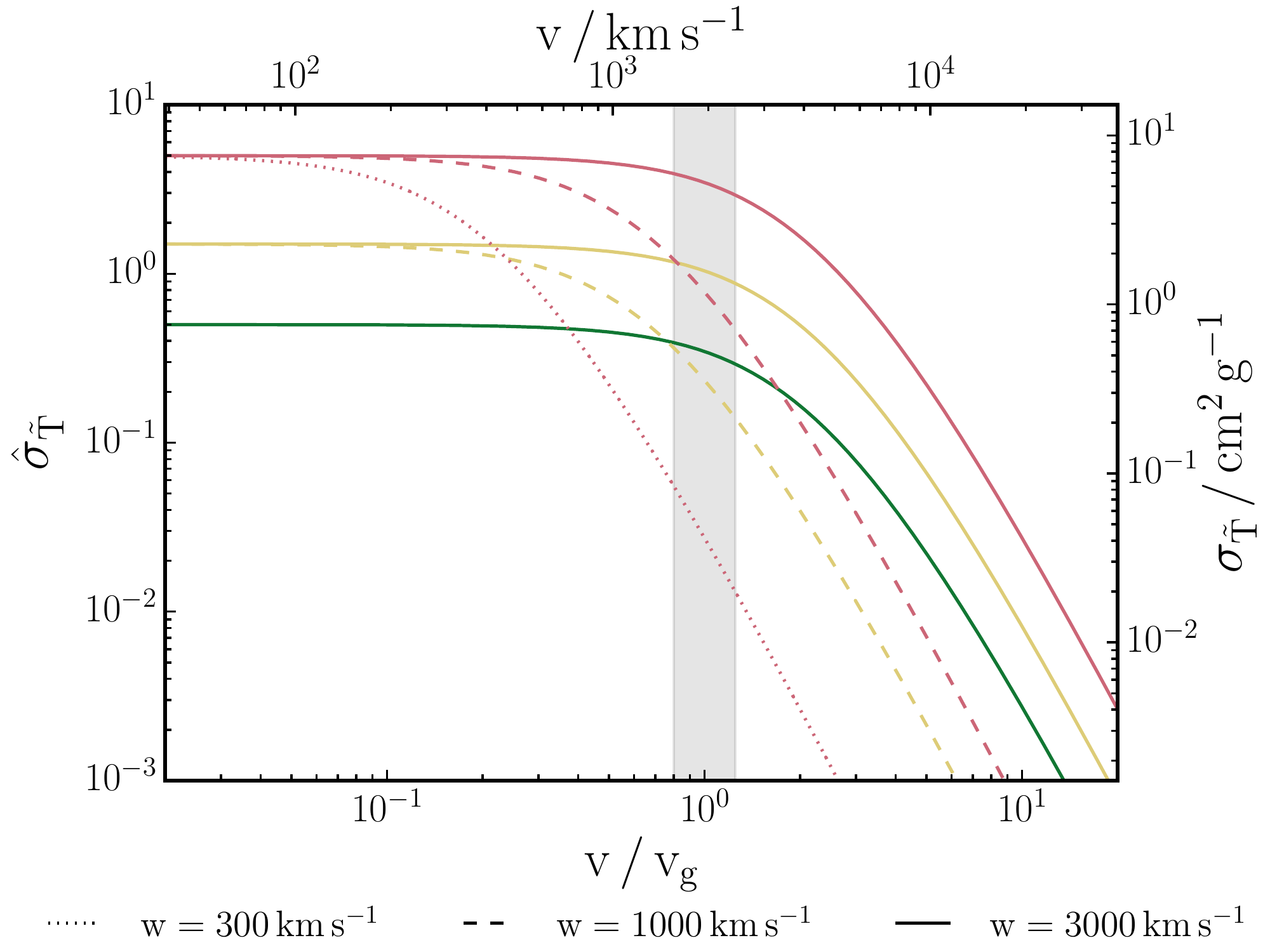}
  	&
  	\includegraphics[width=0.5\textwidth]{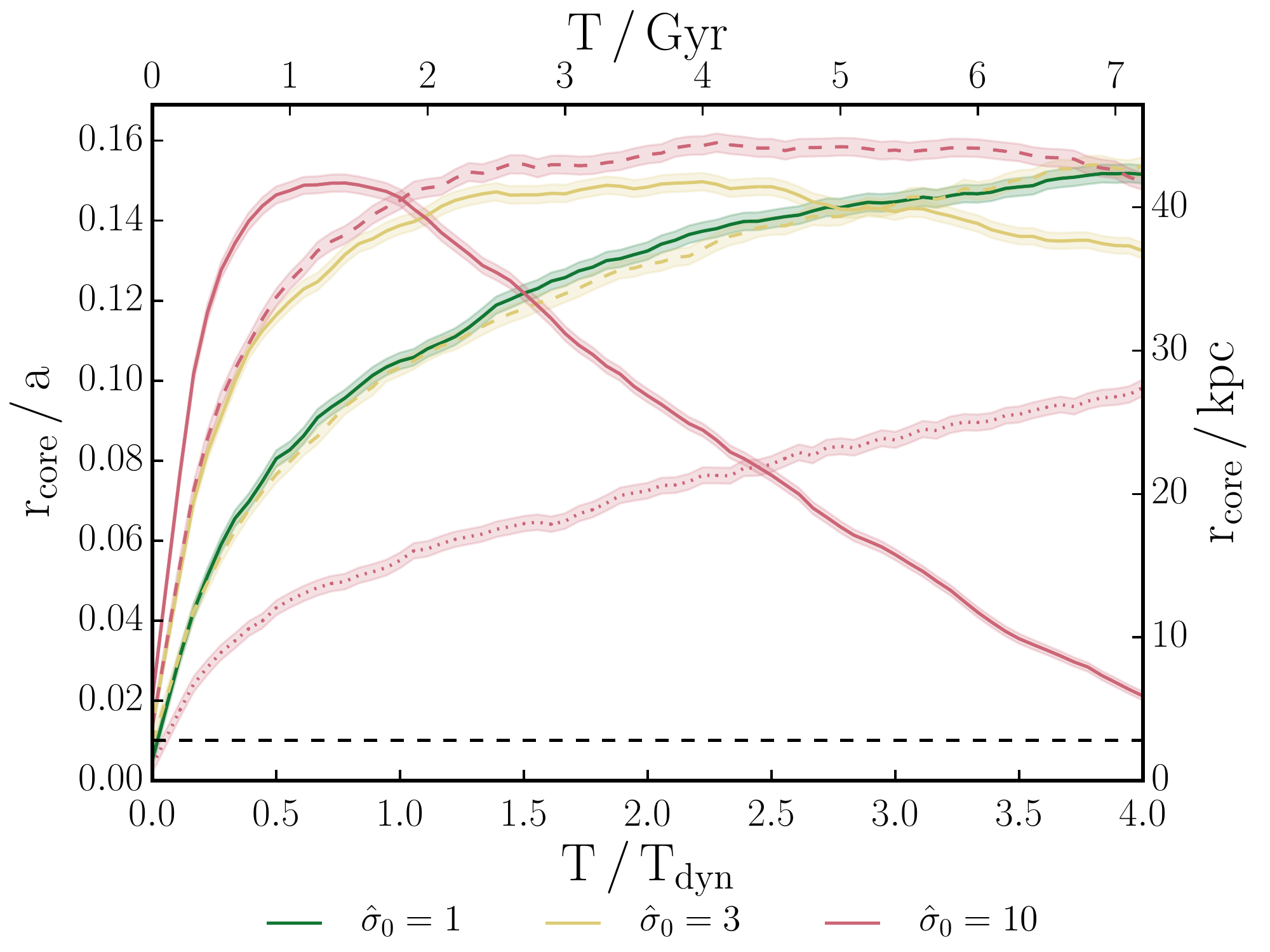}
	 \end{tabular}
	\caption{Left panel: The modified momentum-transfer cross-section as a function of velocity for particles scattering through a Yukawa potential, assuming the Born approximation. The different lines correspond to different values for the DM and mediator particle masses, as well as the coupling strength for their interaction. These three parameters lead to the two astrophysically important parameters that describe the scattering: $\sigma_0$, the cross-section at low velocities when the scattering is isotropic, and $w$, the velocity around which the cross-section transitions from being isotropic and velocity-independent to anisotropic with a cross-section that drops rapidly with increasing velocity. Right panel: the evolution of core size in an isolated Hernquist profile for the different cross-sections shown in the left panel. The evolution of core sizes is approximately captured by the value of $\sigma_{\tilde{T}}$ at $v=v_\mathrm{g}$, where $v_\mathrm{g} \def \sqrt{GM/a}$ is a typical velocity for particles within the halo, marked by the vertical shaded region in the left panel.}
	\label{fig:yukawa}
\end{figure*}

In order to further test the ability of an integrated cross-section to capture the effects of anisotropic scattering, we simulate the same Hernquist profile as in \S\ref{sect:KVI_Hernquist}, this time with scattering from a Yukawa potential assuming the Born approximation. The DM scattering follows equation~\eqref{eq:yukawa_differential_cross-sect_alt} and we simulate cross-sections with three different $w$, with a variety of $\sigma_0$. The modified momentum-transfer cross-sections for the different simulated cross-sections are shown in the left panel of Fig.~\ref{fig:yukawa}.

In the right panel of Fig.~\ref{fig:yukawa} we show how the core sizes evolve for the different particle models shown in the left panel. As the Hernquist halo in question has a typical velocity for particles moving within the halo of $v_g = \sqrt{GM/a} \approx 1950 \kms$, most scattering for the $w=3000 \kms$ models is in the isotropic regime, and the core evolution is similar to that seen in Fig.~\ref{fig:core_radius_evolution} with isotropic scattering. For all of the particle models, the evolution of the core-size is approximately determined by the value of $\sigTt(v_g)$. For example the $\hat{\sigma}_0=3$,  $w=1000 \kms$ cross-section and $\hat{\sigma}_0=1$, $w=3000 \kms$ cross-section show similar evolution of $r_\mathrm{core}$, while having similar values of $\sigTt(v_g)$.

Having simulated these particle models using the full differential cross-section, we can also test both $\sigT$-match and $\sigTt$-match. We simulated each of the particle models from Fig.~\ref{fig:yukawa}, using isotropic scattering and the appropriate $\sigma(v)$ for $\sigT$-match and $\sigTt$-match. Our results show that the momentum transfer cross-section, $\sigma_T$, is not a good quantity to use to match an isotropic cross-section to an anisotropic one. Instead we found that when cross-sections with different angular dependencies are matched by $\sigTt$, the rate of core formation is very similar.

\begin{figure}
        \centering
        \includegraphics[width=\columnwidth]{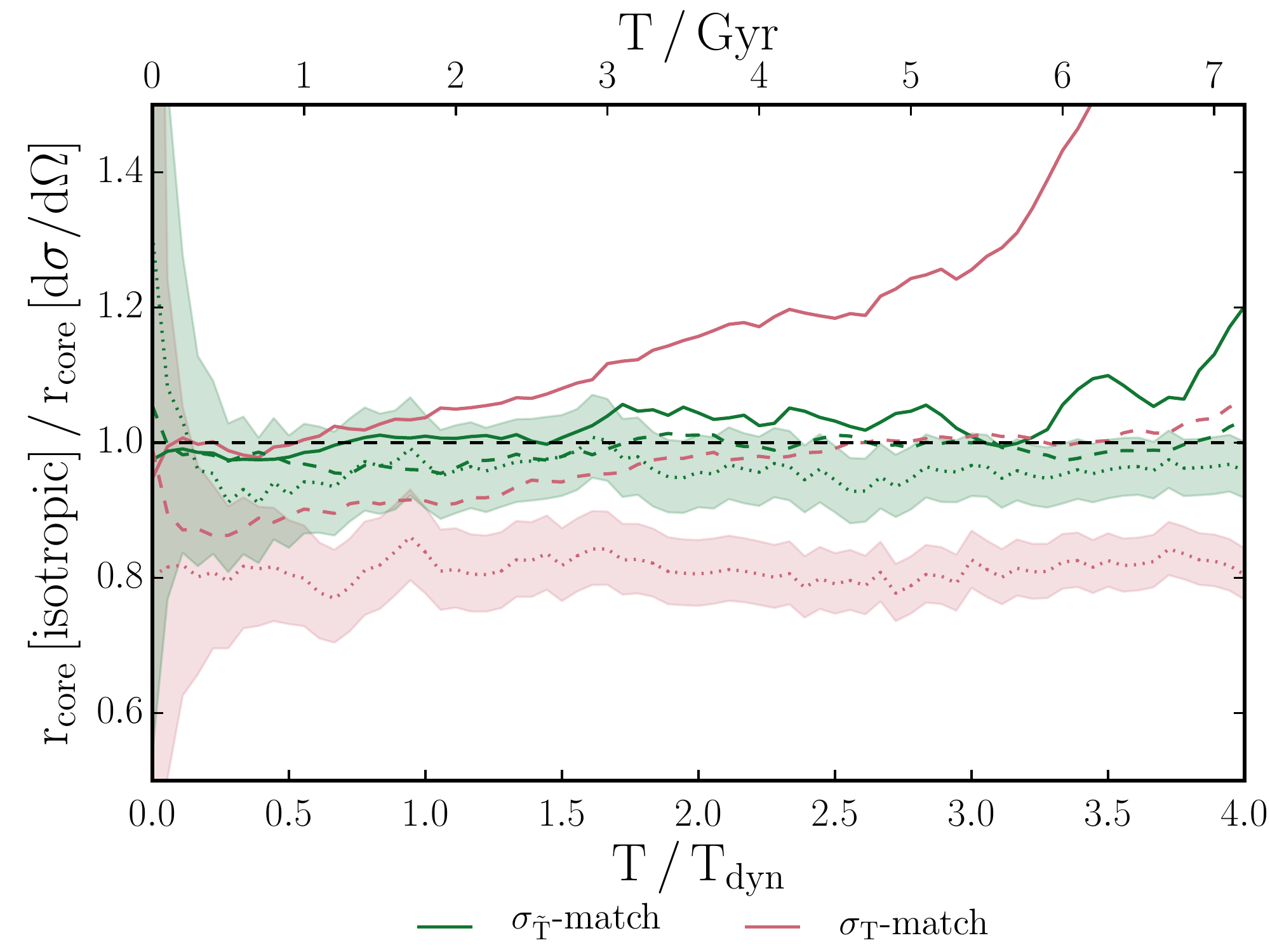}
	\caption{The core size using $\sigT$ or $\sigTt$-match, normalised by the core size from faithfully simulating the underlying particle model, for the three cross-sections with $\hat{\sigma}=10$ from Fig.~\ref{fig:yukawa}. The line styles represent the value of $w$, in the same manner as Fig.~\ref{fig:yukawa}, and the shaded regions around the $w = 300 \kms$ lines are representative of the errors on this ratio -- coming from the widths of the posterior distributions for $r_\mathrm{core}$. $\sigTt$-match works well in all cases, while $\sigT$-match incorrectly predicts the effects of anisotropic scattering on an isolated halo.}
	\label{fig:core_radius_evolution_yukawa_comparison}
\end{figure}

Fig.~\ref{fig:core_radius_evolution_yukawa_comparison} shows the core size from the $\sigT$-match and $\sigTt$-match simulations, divided by the core size from the simulation with the full differential cross-section. For clarity we show only the $\hat{\sigma}_0 = 10$ cross-sections, but found that $\sigT$-match systematically underpredicts the rate of scattering for all of the simulated cross-sections. For the examples shown in Fig.~\ref{fig:core_radius_evolution_yukawa_comparison}, this underprediction in scattering rates with $\sigT$-match manifests itself in core sizes being smaller than in the full differential cross-section simulations with $w = 300 \kms$, and larger when $w = 3000 \kms$. This change in behaviour is because with $w = 3000 \kms$ the halo undergoes core-collapse, and so a lower rate of scattering leads to larger cores at fixed time.  In contrast, $\sigTt$-match correctly predicts the evolution of the core size for all of our simulated cross-sections. At early times the ratio of core sizes using $\sigTt$-match is not unity, however this is because the small cores lead to large errors on this ratio. This applies also to the $\hat{\sigma}_0 = 10$ model at late times, when $r_\mathrm{core}$ is again small. We emphasise that $\sigT$-match has been used in previous work, that has incorrectly estimated the effects of DM models with anisotropic particle scattering \citep[e.g.][]{2012MNRAS.423.3740V,2013MNRAS.431L..20Z,2016PhRvD..93l3527C}.

That $\sigTt$-match leads to an enhanced rate of scattering compared with $\sigT$-match can be understood from considering what happens when scattering is anisotropic, with a large fraction of scattering being by small angles. In this case, $\sigT$ and $\sigTt$ will be similar, as they only differ in how they treat scattering by angles $\theta > \pi/2$. However, for isotropic scattering, $\sigTt = \sigT / 2$, so matching with $\sigTt$ will lead to twice the rate of scattering as matching with $\sigT$. Fig.~\ref{fig:core_radius_evolution_yukawa_comparison} demonstrates that $\sigTt$ more accurately captures the effects of scattering by anisotropic cross-sections. This should be expected given that two particle models could have different $\sigT$, while having indistinguishable particle interactions, purely due to how particles are labelled. 

\section{DM-galaxy offsets in the Bullet Cluster}
\label{sec:offsets_in_BC}

Having demonstrated that the effects in an isolated halo of SIDM with an anisotropic cross-section can be understood by considering the \sTt, we now go on to investigate whether this is still the case in a system with strong directionality. The system we use is based on the merging galaxy cluster 1E~0657-56 (the Bullet Cluster), using the fiducial mass model from \citet{Robertson11022017}. The initial conditions used for all the simulations contain two Hernquist profiles, separated by $4\mpc$, and with a relative velocity of $2970 \kms$ along the line joining the two cluster centres. The main halo and bullet halo have Hernquist density profiles with masses and scale radii $M = \num{3.85e15} \msun$, $a=1290 \kpc$, and $M = \num{2.46e14} \msun$, $a=279 \kpc$ respectively. The mass within each halo is 99\% DM, and 1\% stars, though we use an equal number of DM and star particles ($10^7$ of each). The star particles are distributed as a smooth halo following the DM density profile.

The position estimates for the DM and galaxies were performed following the method described in \S$3.2$ of \citet{Robertson11022017}. This involves simultaneously fitting parametric models for the two haloes to the projected surface density, modelling each halo with a Pseudo Isothermal Elliptical Mass Distribution (PIEMD), which has a 3D density profile
\begin{equation}
\rho(r) = \frac{\rho_0}{(1+r^2 / r_\mathrm{core}^2)(1+r^2 / r_\mathrm{cut}^2)}; \quad r_\mathrm{cut} > r_\mathrm{core}.
\label{PIEMD_density}
\end{equation}
We choose to fit parametric models because it is often done observationally \citep{2005MNRAS.359..417S,2010MNRAS.402L..44R,2012ApJ...757....2G,2015Sci...347.1462H,2015MNRAS.449.3393M,2016ApJ...820...43S} and because fitting two parametric models simultaneously accounts for the main halo when trying to fit the position of the bullet halo. This is an advantage over local position estimates such as shrinking circles -- the 2D analogue of the shrinking spheres approach described in \citet{2003MNRAS.338...14P} -- where the density gradient from the main halo can lead to spuriously large measured offsets between the DM and galaxies in the bullet halo \citep{Robertson11022017}.

\subsection{KVI scattering}
\label{KVI Bullet Cluster}

\begin{figure}
        \centering
        \includegraphics[width=\columnwidth]{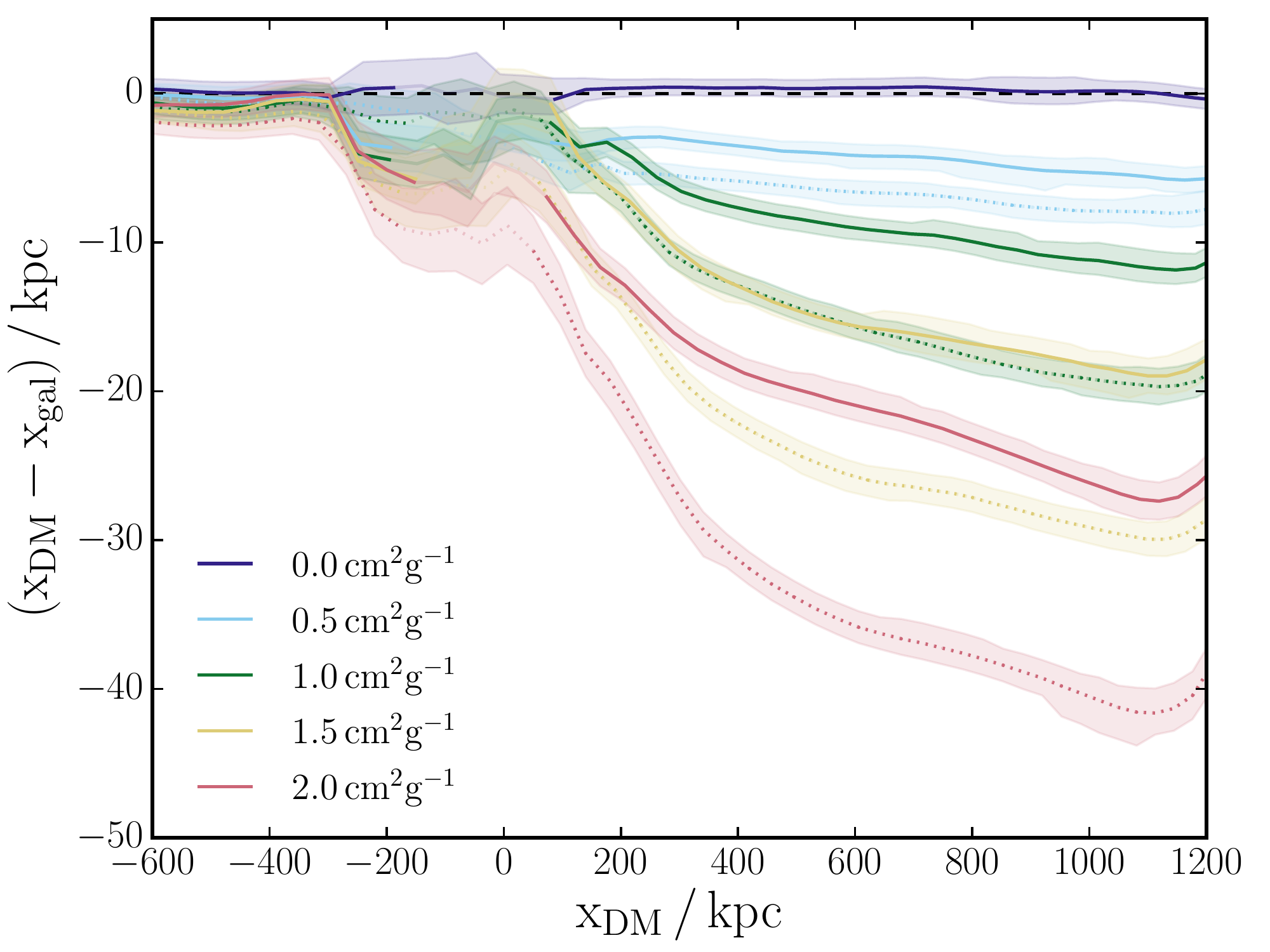}
	\caption{The DM-galaxy offset in the bullet halo of the Bullet Cluster as a function of the bullet halo position (in the centre of mass frame). The solid lines are for isotropic scattering with a cross-section as given in the legend. The dotted lines are for scattering with a KVI cross-section that has the same $\sigTt$ as the corresponding isotropic cross-section. The DM and galaxy positions were determined by fitting two parametric model haloes to the respective projected densities. The separations were calculated every $10 \myr$, and are plotted using a $50 \myr$ moving average. Lines are shown as faint around core passage because measurements of the best-fit halo positions become noisy. In the observed Bullet Cluster the two haloes are separated by $\sim 720 \kpc$, which happens at $\mathrm{x_{DM}} \approx 600 \kpc$ with our mass model.}
	\label{fig:separations_KVI_iso_ePIEMD}
\end{figure}

In Fig.~\ref{fig:separations_KVI_iso_ePIEMD} we show that with KVI scattering the measured offsets between DM and galaxies are $\sim 50\%$ larger than for the isotropic cross-section to which they are matched. This matching was done using $\sigTt$-match, which as demonstrated in Fig.~\ref{fig:core_radius_evolution} leads to core formation rates in isolated haloes that are very similar for isotropic scattering and a matched KVI cross-section. This $\sim 50\%$ increase in offsets was seen throughout the evolution of the merger for each of the four cross-sections simulated: 0.5, 1, 1.5 and 2 $\cmsg$. This result supports the findings of K14, who found that $\sigTt$ was not enough to fully characterise the effects of DM scattering.

\begin{figure*}
        \centering
        \includegraphics[width=\textwidth]{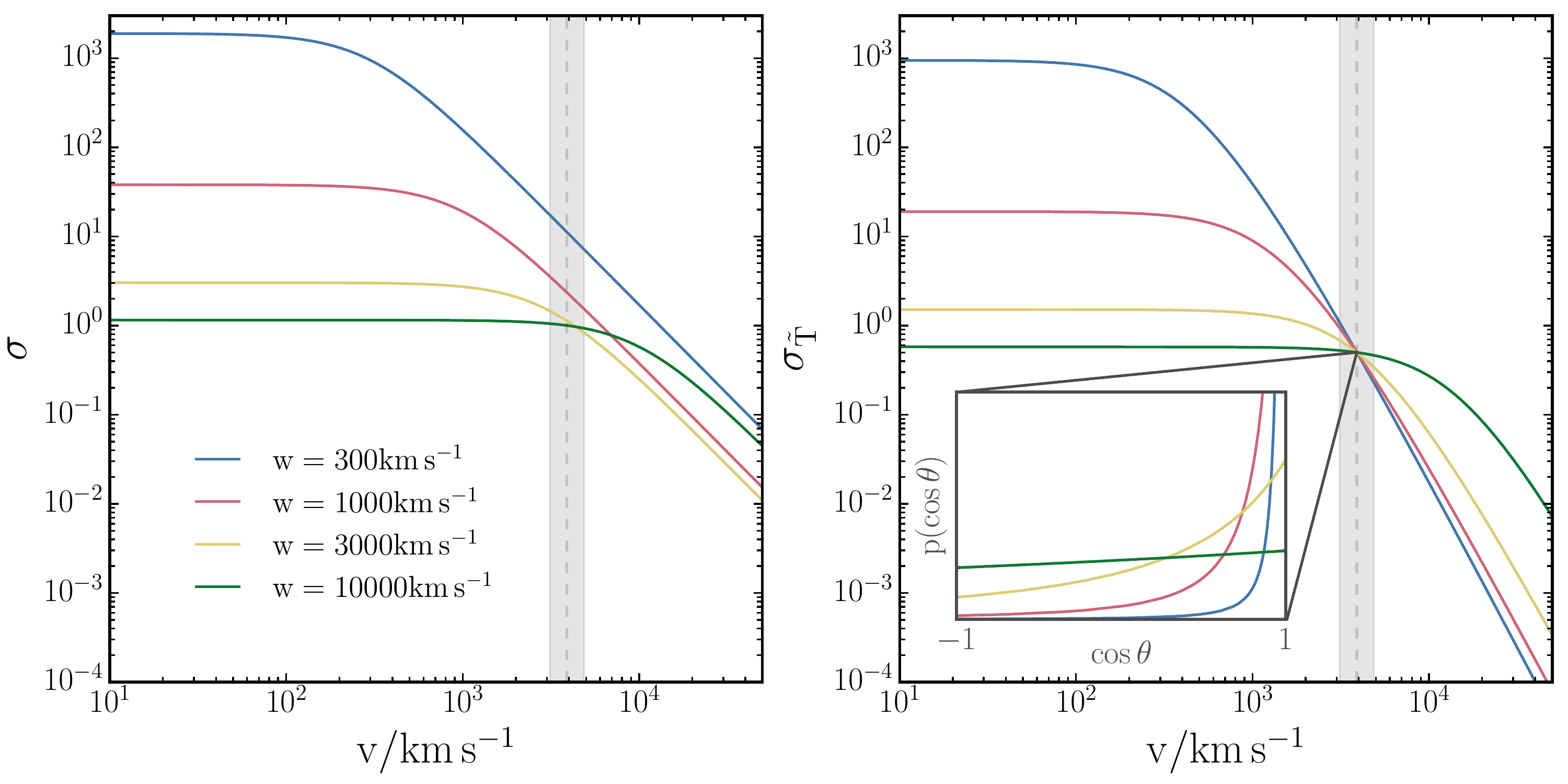}
	\caption{The total cross-section (left panel) and \sTt \, (right panel) for four differential cross-sections. These four cross-sections correspond to the differential cross-section in equation \eqref{eq:yukawa_differential_cross-sect_alt}, with four different values of $w$. The normalisation of the cross-section, $\sigma_0$, was chosen such that the \sTt \, at a velocity of $3900 \kms$ (the relative velocity between the two DM haloes during core passage in our Bullet Cluster simulations) was $0.5 \cmsg$, the same as for isotropic scattering with $\sigma/m = 1 \cmsg$. Inset in the right panel is a plot showing the angular dependence of the four different cross-sections at a velocity of $3900 \kms$. The $w = 300 \kms$ line shows highly anisotropic scattering, with a majority of low-$\theta$ scattering events, while for $w = 10\,000 \kms$ the scattering is almost isotropic ($p(\theta) \propto \sin \theta$).}
	\label{fig:yukawa_bullet_cluster_cross-sects}
\end{figure*}

Intuitively this can be understood: for isotropic scattering, only a small fraction of bullet halo DM particles scatter with a particle from the main halo, and those that do are typically ejected from the bullet halo. The unscattered DM is coincident with the collisionless galaxies, and any measured offset is a result of fitting to the wake of scattered particles. This is not the case with anisotropic scattering, where many more particles can scatter, but each receives only a small momentum kick. The DM particles that have received such a kick lag behind the collisionless galaxies, leading to an offset between the galaxies and DM.

The KVI cross-sections were simulated using $\theta_\mathrm{min}=0.025$. Given that Fig.~\ref{fig:separations_KVI_iso_ePIEMD} demonstrates that cross-sections with the same $\sigTt$ can lead to different DM-galaxy offsets, one might worry that the results in Fig.~\ref{fig:separations_KVI_iso_ePIEMD} are dependent on $\theta_\mathrm{min}$. We tested for convergence with respect to $\theta_\mathrm{min}$ by running the KVI cross-section $\sigTt$-matched to isotropic $1 \cmsg$ with $\theta_\mathrm{min}=0.00625$ and $0.1$. The results from these tests were in agreement with each other and the $\theta_\mathrm{min}=0.025$ results.

\subsection{Yukawa-potential scattering}
\label{sect:yukawa_BC}

Having shown that using isotropic scattering to emulate the effects of the KVI anisotropic scattering would lead to an underprediction in the measured offset between DM and galaxies, we now perform similar tests with the more complicated differential cross-section described by equation~\eqref{eq:yukawa_differential_cross-sect_alt}. As well as having a velocity dependence, this cross-section has an angular dependence that changes with velocity. This means that the behaviour of such a particle in galaxy cluster collisions could be very different from in the cores of dwarf galaxies, due to the very different velocity scales and the anisotropic nature of a cluster collision.

\subsubsection{Simulated cross-sections}

We simulate four different cross-sections, with $w= 300, 1000, 3000$ and $10\,000 \kms$. The relative velocity between the two DM haloes in our simulations is $3900 \kms$ at the time of core passage, so this range of $w$ values was chosen to bracket inter-halo scattering in the isotropic regime ($w=10\,000 \kms$) all the way down to Rutherford-like scattering ($w=300 \kms$). With fixed $\sigma_0$ the low-$w$ cross-sections would have much lower $\sigTt$ at $\sim 3900 \kms$ than those with high-$w$. In order to keep the offsets with the different cross-sections measurable, we normalise the different cross-sections such that $\sigTt(v=3900 \kms) = 0.5 \cmsg$, the same $\sigTt$ as isotropic scattering with $\sigma/m = 1 \cmsg$.

The four simulated cross-sections are displayed in Fig.~\ref{fig:yukawa_bullet_cluster_cross-sects}, which shows both $\sigma(v)$ and $\sigTt(v)$, as well as the angular dependence, $p(\theta)$, at $v=3900 \kms$. In the $v>w$ regime $\sigTt$ rises rapidly towards low velocities, such that for these particle models, tests on smaller scales may provide better constraints. Nevertheless, simulations of dwarf galaxies have shown that at velocities $\sim 40 \kms$ the cross-section could be as large as $50 \cmsg$ \citep{2015MNRAS.453...29E} without being in tension with observations of Milky Way or Local Field dwarf galaxies, such that these models may not be as outlandish as they first appear. Even if such large cross-sections cannot be accommodated at low velocities, these differential cross-sections correspond to a perturbative treatment of Yukawa scattering. \citet{2013PhRvD..87k5007T} have shown that quantum mechanical and non-perturbative effects can become important when $\alpha_\chi m_\chi / m_\phi \gtrsim 1$ and $m_\chi v / m_\phi c \lesssim 1$ respectively. In this ``resonant regime'', quasi-bound states in the potential can lead to resonances or antiresonances that could alter the cross-section substantially at low velocities. Using a model such as the $w=300 \kms$ one is therefore interesting as it probes what would happen in a galaxy cluster collision, where the collision speed places inter-halo scatters deep within the anisotropic regime.

\subsubsection{DM-galaxy offsets}

Fig.~\ref{fig:shrinkcirc_X_R_yukawacomparison} shows the measured DM-galaxy offsets for the different Yukawa cross-sections at the time of the observed Bullet Cluster -- defined as the snapshot where the separation between the two haloes is closest to $720 \kpc$. These offsets are calculated using a shrinking circles approach down to different final radii. We stress that the offsets found through shrinking circles can be anomalously large due to a bias that comes from the presence of a nearby halo \citep{Robertson11022017}, and so these offsets should not be compared with observations of DM-galaxy offsets. However, the shrinking circles procedure is used as shrinking to different radii provides insight into the 2D distribution on different scales, and is useful for comparing the effects of different scattering cross-sections. To allow for a comparison with observed offsets we also plot in Fig.~\ref{fig:shrinkcirc_X_R_yukawacomparison} the offset measured by fitting parametric models to the projected mass distribution. For all simulated cross-sections these are less than $10 \kpc$ and decrease with increasing angular dependence.

The largest offset arises when the cross-section is closest to isotropic, which is surprising given that these cross-sections were matched to have the same $\sigTt$ at the collision velocity of the two DM haloes and in Fig.~\ref{fig:separations_KVI_iso_ePIEMD} we demonstrated that the more anisotropic scattering cross-section (KVI) lead to larger DM-galaxy offsets than the $\sigTt$-matched isotropic cross-section. To investigate this apparent discrepancy further, we now isolate the effects of angular and velocity dependence by running  $\sigTt$-matched isotropic versions of our Yukawa cross-sections, as well as Yukawa cross-sections with the velocity dependence removed.

\subsubsection{$\sigTt$-matched Yukawa scattering}

\begin{figure}
        \centering
        \includegraphics[width=\columnwidth]{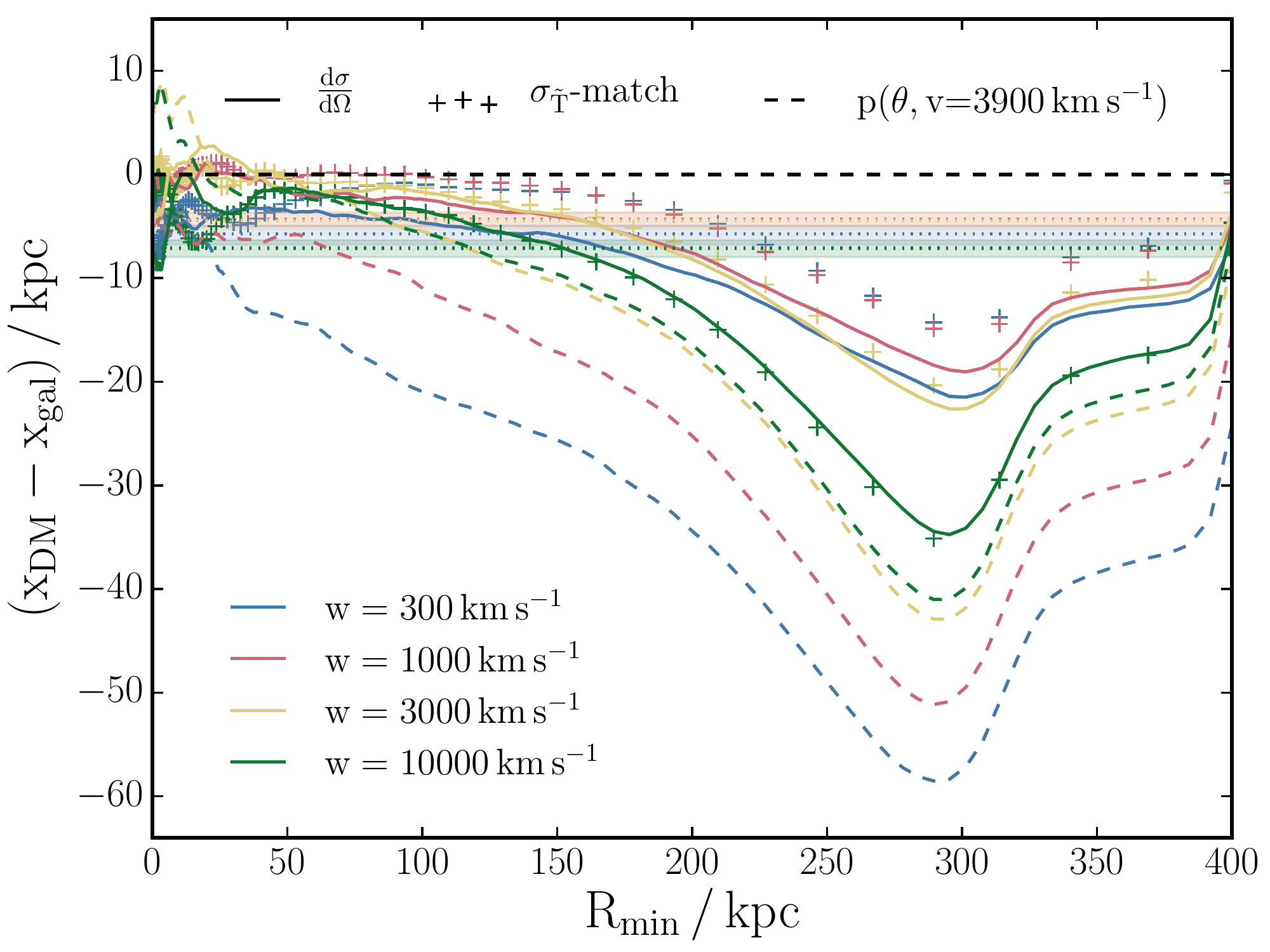}
	\caption{The shrinking circles offsets for different simulated particle physics models, showing how the DM-galaxy separation varies as a function of the final radius to which the circles are shrunk. The solid lines show the results using the full differential cross-section for Yukawa scattering under the Born approximation, while the crosses show the results of trying to mimic this scattering using suitably matched isotropic scattering (matched using $\sigTt$). The dashed lines show what happens when we remove the velocity dependence of the Yukawa scattering models, by using the Yukawa differential cross-section at $v=3900 \kms$ at all velocities. Finally, the horizontal bands represent the DM-galaxy offsets for the full differential cross-section cases, as measured by fitting parametric models to the projected surface density.}
	\label{fig:shrinkcirc_X_R_yukawacomparison}
\end{figure}

In an isolated halo, Fig.~\ref{fig:core_radius_evolution_yukawa_comparison} demonstrated that using $\sigTt$-match allows us to use velocity-dependent isotropic scattering to predict the effects of Yukawa scattering. To test whether this still works in a system with a strong directionality, we plot the DM-galaxy offsets when using $\sigTt$-match in Fig.~\ref{fig:shrinkcirc_X_R_yukawacomparison}. For the high-$w$ cross-sections (where scattering at $v \approx 3900 \kms$ is fairly isotropic anyway) this procedure is effective and the results are similar to those from using the full differential cross-section. For the low-$w$ cross-sections $\sigTt$-match underpredicts the separations from using the full differential cross-section.

\subsubsection{The effects of velocity-dependent scattering on DM-galaxy offsets}

The DM-galaxy offsets with $\sigTt$-matched Yukawa scattering are in agreement with our findings in \S\ref{KVI Bullet Cluster} that isotropic scattering leads to smaller offsets than anisotropic scattering when the cross-sections have the same $\sigTt$. However, they leave the question of why the most isotropic cross-section ($w=10\,000 \kms$) leads to offsets that are substantially larger than for the more anisotropic cases. There are two primary reasons for this, both related to the velocity dependence of the anisotropic cross-sections:
\begin{enumerate}
 \item While the relative velocity between the centres of mass of the two DM haloes is $\sim 3900 \kms$ during core passage, the velocity of particles within their own haloes transverse to the collision axis means that for inter-halo pairs of particles the mean pairwise velocity is larger than $3900 \kms$. Assuming isotropic velocity dispersions in the two haloes, with 1D velocity dispersions of $1200 \kms$ and $600 \kms$ for the main and bullet halo respectively leads to an average pairwise velocity of $\sim 4350 \kms$. At this velocity, the most anisotropic cross-sections, which have the steepest $\sigTt(v)$, have the lowest $\sigTt$ as they were normalised to have the same $\sigTt$ at a lower velocity, $3900 \kms$. 
  \item The steep gradient of $\sigma(v)$ around $v = 3900 \kms$ in the low-$w$ models means that pairs of particles with low pairwise velocities are significantly more likely to scatter than those with high pairwise velocities. This means that of the particles in the bullet halo, it is those moving in the opposite direction to the motion of the bullet through the main halo that are most likely to scatter with a particle from the main halo. Preferentially scattering those particles travelling backwards over those travelling forwards leads to a forward shift in the position of the DM halo compared to all particles scattering with equal probability. This in turn reduces the DM-galaxy separation.
\end{enumerate}

\begin{figure}
        \centering
        \includegraphics[width=\columnwidth]{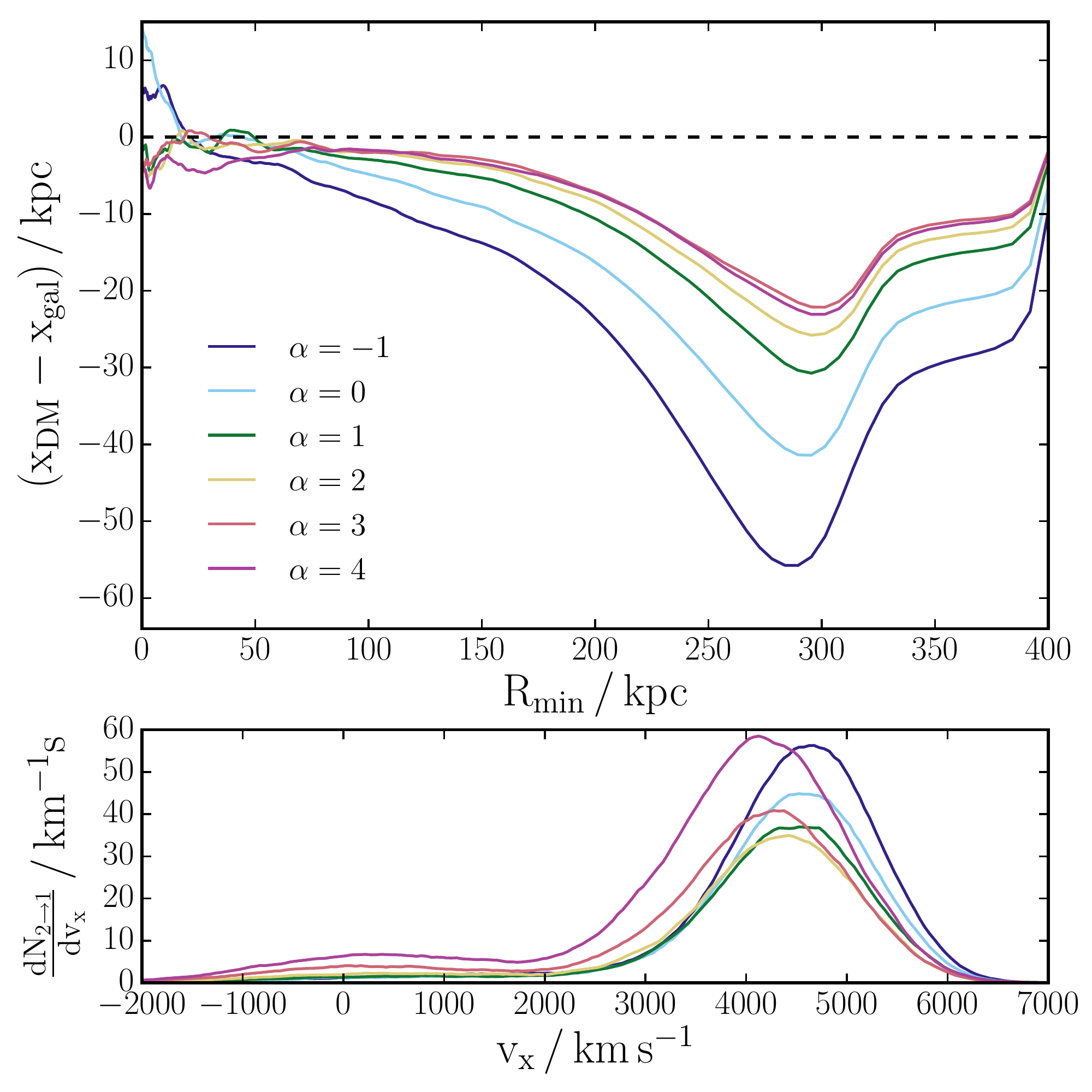}
	\caption{Top panel: The DM-galaxy offset at the time of the observed Bullet Cluster with isotropic DM scattering with a power-law velocity-dependence: $\sigma(v)/m = \left( v / 4350 \kms \right)^{-\alpha} \cmsg$. The cross-section with the strongest velocity-dependence ($\alpha = 4$) has a smaller offset than the velocity-independent case ($\alpha=0$), despite the former having a greater fraction of bullet halo particles that scatter from particles in the main halo. Bottom panel: The distribution of $v_x$, the velocity along the collision axis, for particles from the bullet halo before they scatter with a particle from the main halo. As well as changing the total number of scattered particles, increasing $\alpha$ shifts the distribution towards lower $v_x$.}
	\label{fig:shrinkcirc_X_R_powerlaw_v-dep}
\end{figure}

The second reason is elucidated in Fig.~\ref{fig:shrinkcirc_X_R_powerlaw_v-dep} where we show the DM-galaxy offsets with isotropic scattering and power-law velocity-dependent cross-sections. We simulate cross-sections of the form
\begin{equation}
\frac{\sigma(v)}{m} = \left( \frac{v}{4350 \kms} \right)^{-\alpha} \cmsg,
\label{v-dep_cross-sections}
\end{equation}
such that all cross-sections have $\sigma/m = 1 \cmsg$ at the average pairwise velocity for particles drawn randomly from the two different haloes. To cut-off the low velocity divergence in the cross-section, we capped the cross-sections from equation \eqref{v-dep_cross-sections} to $100 \cmsg$.

We find that despite having the highest rate of inter-halo scattering, the $\alpha=4$ case also has the lowest DM-galaxy offsets, even though scattering is isotropic in all cases. This is explained by the selection effect of a velocity dependent cross-section, such that those particles that scatter preferentially had certain properties (see bottom panel of Fig.~\ref{fig:shrinkcirc_X_R_powerlaw_v-dep}). Using terminology whereby the bullet halo moves to the `right': not just are bullet halo particles that scatter more likely to have been moving left relative to the bullet halo, the particles from the main halo with which they scatter are likely to be moving right relative to the main halo. This means that with large $\alpha$ the majority of scatters take place with a relative velocity lower than the mean pairwise velocity, and so transfer less momentum between the two haloes. Also, if scattered particles are ejected from the bullet halo, and these scattered particles were preferentially moving left, the remaining DM particles will preferentially be moving right with respect to the bullet halo, which pushes the measured position of the DM halo right and reduces the DM-galaxy offset.  

\subsubsection{Isolating Yukawa scattering's angular-dependence}

Returning to Fig.~\ref{fig:shrinkcirc_X_R_yukawacomparison}, we isolate the effects of the angular-dependence of Yukawa scattering at different $v/w$ by looking at the DM-galaxy separations in the Bullet Cluster including DM scattering that uses the angular dependence and normalisation of our Yukawa models at $v=3900 \kms$ at all velocities. For these cases we recover our previous result that more anisotropic scattering, in this case lower $w$, results in larger offsets at fixed $\sigTt$. The smaller offsets with smaller $w$ seen for the full differential cross-section runs do not contradict our findings in \S\ref{KVI Bullet Cluster}. Rather, the effects of a steeply decreasing $\sigTt(v)$ more than compensate for the increased angular dependence.

\section{Conclusions}
\label{sec:conclusions}

We have explored simulations of SIDM where the scattering is anisotropic. Anisotropic cross-sections arise when the rate of scattering depends on the amount of exchanged momentum, and are natural in models with a velocity-dependent DM scattering cross-section. We considered two different models of anisotropic scattering, one without any velocity-dependence (KVI) and one that has a total cross-section and angular dependence that varies with velocity and corresponds to Yukawa scattering under the Born approximation.

For both of these anisotropic models the evolution of an isolated halo could be adequately captured by treating the scattering as isotropic (Fig.~\ref{fig:core_radius_evolution} and Fig.~\ref{fig:core_radius_evolution_yukawa_comparison}), provided that the isotropic cross-section is suitably matched to the underlying model. We find that what needs to be matched between different cross-sections in order for them to behave in a similar way is $\sigTt$, defined in equation \eqref{eq:sigTt}. This is similar to the momentum transfer cross-section, $\sigT$, that has been used by previous authors  \citep{2012MNRAS.423.3740V,2013MNRAS.430.1722V,2013MNRAS.431L..20Z,2014MNRAS.444.3684V} to match an underlying particle physics model on to a velocity-dependent but isotropic scattering cross-section that is more easily simulated. For cross-sections that are close to isotropic the matching scheme chosen is not particularly important, but when the scattering is highly anisotropic (with the majority of particles scattering by $\theta \ll \pi$) there is a factor of 2 difference in the $\sigT$-matched and $\sigTt$-matched isotropic cross-sections. This is because $\sigT$ overestimates the ability of isotropic scattering to alter dynamics, because scattering by large angles ($\sim \pi$) leads to a large amount of momentum transfer, despite leaving the system relatively unchanged (the two particles have simply switched places). In Fig.~\ref{fig:core_radius_evolution_yukawa_comparison} we demonstrate that this results in $\sigT$-matched isotropic scattering underpredicting the effects of an anisotropic cross-section, with cores using $\sigT$-matched isotropic scattering evolving slower than using $\sigTt$-matched isotropic scattering, which in turn agreed with the results of using the full anisotropic cross-section.

We went on to investigate how the $\sigTt$ matching scheme works in a system that has a strong directionality, namely Bullet Cluster-like galaxy cluster collisions. With an anisotropic but velocity-independent cross-section, we found that the distribution of DM was not correctly captured by using matched isotropic scattering, which underpredicted the size of DM-galaxy offsets induced by KVI scattering by $\sim 33\%$. For the case of Yukawa scattering in a galaxy cluster collision we found that the strong velocity dependence of the cross-section in regimes where the cross-section is anisotropic, leads to a suppression of the DM-galaxy offsets. Using matched isotropic scattering still underpredicts the DM-galaxy offset (crosses in Fig.~\ref{fig:shrinkcirc_X_R_powerlaw_v-dep}), but these offsets are small anyway due to the velocity dependence. This suppression of DM-galaxy offsets is not simply because velocity dependent cross-sections must be small at typical galaxy cluster velocities to be reasonable at lower velocities. In fact, the small offsets result even when the velocity dependent cross-sections are boosted to have a substantial $\sigTt$ at cluster velocities. The small offsets are a result of the gradient in $\sigma(v)$, which results in particle pairs with low relative velocities being more likely to scatter than others. These low velocity pairs are made of particles that move within their halo in the opposite direction to the bulk velocity of their halo, and preferentially scattering these particles leaves a population of unscattered particles moving faster than the bulk velocity of the halo. This shifts the measured DM position forwards reducing any DM-galaxy offset.

Ignoring the angular-dependence of SIDM models and instead using suitably matched isotropic cross-sections appears to work well in isotropic systems such as an isolated halo, but can lead to differences from the true result in anisotropic systems. Despite these differences, merging galaxy clusters do not appear to be a good place to constrain Yukawa-like DM scattering, as the cross-section at cluster velocities would be lower than in smaller objects, and the increased DM-galaxy separation due to the anisotropic nature of the scattering is more than compensated for by the decreased DM-galaxy separation coming from the gradient in $\sigma(v)$ about the collision velocity of the clusters. Previous results that have simulated an anisotropic scattering model using appropriately matched isotropic scattering have typically focused on the density profiles of dwarf galaxies. Our results in isolated haloes suggest these results are probably robust to changing from isotropic scattering to using the underlying differential cross-section. That being said, a cosmologically formed DM halo evolves through numerous mergers, and it is unclear if incorrectly modelling the effects of SIDM in these mergers could lead to differences in the final density profile. This will need to be addressed in the future by including anisotropic scattering in cosmological simulations.

\vspace{-2em}
\section*{Acknowledgements}
\vspace{-0.5em}
This work would have not be possible without Lydia Heck and John Helly's technical support and expertise. We also thank Richard Bower, Matthieu Schaller, David Harvey, Felix Kahlhoefer, and Stacy Kim for discussions about simulating SIDM, and the anonymous referee for suggestions that helped to improve this paper.

This work was supported by the Science and Technology Facilities Council grant numbers ST/K501979/1 and ST/L00075X/1. RM was supported by the Royal Society. This work used the DiRAC Data Centric system at Durham University, operated by the Institute for Computational Cosmology on behalf of the STFC DiRAC HPC Facility (www.dirac.ac.uk). This equipment was funded by BIS National E-infrastructure capital grant ST/K00042X/1, STFC capital grants ST/H008519/1 and ST/K00087X/1, STFC DiRAC Operations grant  ST/K003267/1 and Durham University. DiRAC is part of the National E-Infrastructure.
\vspace{-2em}

\bibliographystyle{mnras}

\bibliography{MS}

\appendix

\section{Implementation and testing of DM scattering}
\label{App:scattering_implementation}

\vspace{-0.5em}
\subsection{Implementation of anisotropic scattering}
\label{implementation_of_ang_dep}
\vspace{-0.5em}

From considerations of the solid angle at different polar angles, the probability density function for scattering by an angle $\theta$ is
\begin{equation}
\label{eq:p_theta}
p(\theta) = \frac{2 \pi \sin \theta}{\sigma} \frac{\D \sigma}{\D \Omega}.
\end{equation}
Integrating this, we get the cumulative distribution function, 
\begin{equation}
\label{eq:P_theta}
P(\theta) = \int_0^\theta p(\theta') \, \D \theta',
\end{equation}
which is the probability that a particle scatters by an angle less than $\theta$.

For particles due to scatter, a polar scattering angle can be drawn from $p(\theta)$ as the $\theta$ that satisfies
\begin{equation}
\label{eq:P_theta_X}
P(\theta) = X
\end{equation}
where $X$ is a random variable with a uniform distribution in the interval $[0,1]$.

For a general differential cross-section, the inverse of $P(\theta)$ is not necessarily analytical. To allow us to simulate cross-sections with general angular dependence, we numerically find solutions to equation \eqref{eq:P_theta_X} at $N_\theta$ values of $X$ distributed uniformly in the interval $[0,1]$. The $X_i$ take the values $X_i = \frac{i-1/2}{N_\theta}$ where $i = \{1,2,3, ... ,N_\theta \}$, and we label the angles uniformly drawn from $p(\theta)$, $\theta_i$ (i.e. $P(\theta_i) = X_i$). For two particles that scatter, finding a polar scattering angle is then just a case of drawing an integer $i$ from the interval $[1,N_\theta]$ and setting $\theta = \theta_i$.

\vspace{-0.5em}
\subsection{Implementation of velocity-dependant angular-dependence}
\label{sec:implementation_of_arbitrary_cross-sect}
\vspace{-0.5em}

In general, the angular and velocity dependence of a scattering cross-section need not be separable, and $P(\theta)$ can vary with velocity. For these cases, the discussion in Appendix \ref{implementation_of_ang_dep} can be easily extended by generating a set of $\theta_i$ for each of $N_\mathrm{v}$ velocities, where $N_\mathrm{v}$ must be large enough that $p(\theta)$ does not vary substantially from $v_i$ to $v_{i+1}$. Using this as well as a velocity-dependent $\sigma(v)$ in equation \eqref{eq:P_ij} allows us to simulate particle scattering with a general differential cross-section.

Throughout this work, we used $N_\theta, N_\mathrm{v} = 1000$ with the $v_i$ logarithmically spaced from 0.01 to $10\,000 \kms$. At velocities below $0.01 \kms$ the cross-section and angular-dependence were set as if $v= 0.01 \kms$, while at velocities greater than $10\,000 \kms$ the cross-section was set to zero. This was to reflect the fact that for Yukawa-like models the cross-section does not vary at low velocities, and falls off rapidly at high velocities. The values of $N_\theta$ and $N_\mathrm{v}$ could be increased if required, as could the range of velocities covered, but these values were found to be sufficient for the cross-sections and systems we simulated here.

\vspace{-0.5em}
\subsection{Testing generalised scattering}
\label{sec:testing_ang_dep}
\vspace{-0.5em}

To test our implementation of SIDM with velocity and angular dependent cross-sections we ran test cases with a cube of particles moving through a uniform slab of stationary particles. Particles in the cube all moved with a common velocity $v_\mathrm{cube}$ through the slab and there were no gravitational forces. We used the differential cross-section for Yukawa scattering described by equation \eqref{eq:yukawa_differential_cross-sect_alt} and ran the test at five different $v_\mathrm{cube}$, ranging from $0.1 w$ to $10w$. The cross-section normalisation $\sigma_0$ and the projected density of slab particles were chosen such that 10\% of the $N_\mathrm{cube}=10^6$ cube particles would be scattered if the scattering was in the isotropic regime ($v \ll w$) and particles were not allowed to scatter more than once.

\begin{figure}
        \centering
        \includegraphics[width=\columnwidth]{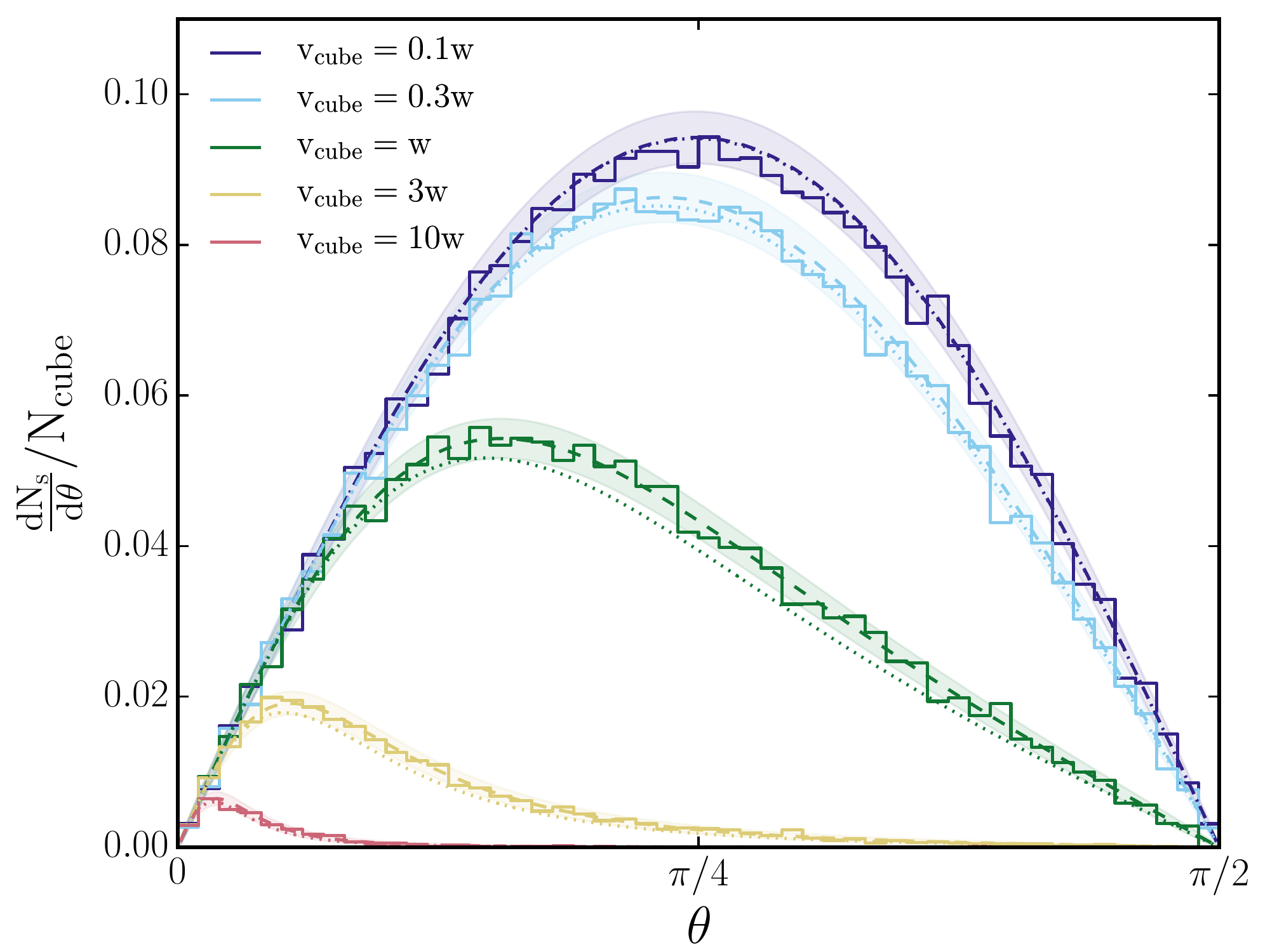}
	\caption{The number of particles that scatter by different polar angles for a cube of DM particles moving through a uniform slab of particles at a speed $v_\mathrm{cube}$. The scattering cross-section was Yukawa scattering under the Born approximation - see equation \eqref{eq:yukawa_differential_cross-sect_alt}. Different line colours correspond to different $v_\mathrm{cube}$, which changes the normalisation and angular dependence of the Yukawa cross-section. The solid lines show scatters into different bins of angle measured in our test simulations, while the dashed lines and shaded regions show the analytically predicted distribution of scattering angles for each of these simulations and the expected $2 \sigma$ Poisson variation. The dotted lines show the prediction for a cube velocity of $(v_{i+5} / v_i) v_\mathrm{cube}$ and are described further in Appendix \ref{sec:testing_ang_dep}. With increasing $v_\mathrm{cube}$ the number of scatters drops, and the scattering becomes more anisotropic.}
	\label{fig:yukawa_scattered_distributions}
\end{figure}

In Fig.~\ref{fig:yukawa_scattered_distributions} we show the results of these test cases, plotted as the number of particles that scatter, $N_\mathrm{s}$, per unit polar angle. These agree with the predictions that were made using $\frac{\D \sigma}{\D \Omega}(v_\mathrm{cube})$ and the projected density of DM through the slab. To make these predictions, the number of expected scatters was calculated using $\sigma (v_\mathrm{cube})$. Their angular distribution was then calculated by transforming the relevant $p(\theta)$ into the frame of the slab, from the centre of momentum frame of the collisions where it is defined. As well as the predicted distribution at velocity $v_\mathrm{cube}$, we also plot the predicted distribution at $(v_{i+5} / v_i) v_\mathrm{cube}$, where the $v_i$ were determined using $N_\mathrm{v} = 1000$ and velocities in the range 0.01 to $10\,000 \kms$. Changing velocity by only one bin led to imperceptibly small changes in the scattered distribution, and the small change in the predicted distribution when increasing velocity by five velocity-bins justifies our choice of $N_\mathrm{v}$.

\bsp
\label{lastpage}

\end{document}